# Low-temperature thermal Hall conductivity of Pr$_2$Zr$_2$O$_7$ single crystal

*Wenjun Chu[1]\* and Xuefeng Sun[2]*

[1]Hefei National Research Center for Physical Sciences at the Microscale, University of Science and Technology of China, Hefei, Anhui 230026, China

[2]Department of Physics and Key Laboratory of Strongly-Coupled Quantum Matter Physics (CAS), University of Science and Technology of China, Hefei, Anhui 230026, People's Republic of China

ABSTRACT: To probe the peculiar excitations spinons and magnetic monopoles in the quantum spin ice candidate Pr$_2$Zr$_2$O$_7$, we studied the low-temperature thermal Hall conductivity ($\kappa_{xy}$) and thermal conductivity ($\kappa_{xx}$) of Pr$_2$Zr$_2$O$_7$ single crystal with magnetic fields applied along the [111] axis. The magnetic field dependencies of $\kappa_{xx}$ suggest the roles of magnetic excitations in thermal conductivity, that is, the emergent magnetic monopoles can transport heat at $T > 1.4$ K and spinons mainly scatter phonons at lower temperatures. The finite $\kappa_{xy}$ was observed at low fields of several Tesla and was discussed to be related to the magnetic excitations, including magnetic monopoles as well as spinons.

## 1. Introduction

The rare-earth pyrochlore materials have been a focus for the study of the geometrical spin frustration. They exhibit a wealth of exotic magnetic properties, including the classical spin ice for rare-earth ions with local Ising anisotropy, like $Ho_2Ti_2O_7$ and $Dy_2Ti_2O_7$,[1–4] and the so-called quantum spin ice for the still enigmatic $Tb_2Ti_2O_7$.[5] Recently, the pyrochlore material $Pr_2Zr_2O_7$, which contains the non-Kramers ion $Pr^{3+}$, was proposed as a possible quantum spin ice. It has been found to be lacking of long-range order down to very low temperatures along with an unexpected spin fluctuation spectrum.[6-8] As a particular kind of quantum spin liquids, the quantum spin ice has not only spin excitations but also magnetic monopoles.[9]

Thermal transport is one of the main probes for studying the exotic spin excitations in the quantum magnetic materials. It has the advantage of detecting the itinerant excitations and can avoid effects of localized excitations caused by impurities which are often inevitable in these materials.

In addition, further details of magnetic excitations can be studied by using the thermal Hall effect measurement. So far, three kinds of thermal Hall effects have been reported in insulators: thermal Hall effects of magnons in ordered magnets,[10-13] spin excitations in quantum disordered magnets,[14-20] and phonons.[21-26] The thermal Hall effects of spin excitations in paramagnetic states have also been studied theoretically.[27-30] These thermal Hall effects have been reported recently for $Tb_2Ti_2O_7$,[14] the frustrated kagomé system,[15-17] and Kitaev compounds.[18-20] However, it is usually difficult to distinguish the thermal Hall effects from the phonons and magnetic excitations. For example, there are some controversies on the mechanism of thermal Hall effect in $Tb_2Ti_2O_7$ and α-$RuCl_3$. Although the thermal Hall effect in these materials was proposed to originate from spinons, the role of phonons cannot be clearly ruled out.[14,18-20,31]

In this work, we study the thermal-transport measurements of $Pr_2Zr_2O_7$ single crystals. It is found that $Pr_2Zr_2O_7$ is a prominent frustrated magnet in which the spin liquid state shows thermal Hall effects of both spinons and monopoles in low fields.

## 2. Experiment and methods

High-quality $Pr_2Zr_2O_7$ single crystal were grown by using a floating-zone technique. The sample for thermal conductivity measurements is cut precisely along the crystallographic axes with typical dimension of 3.06 × 1.37 × 0.151 $mm^3$ after being oriented by using the x-ray Laue photographs. The thermal conductivity and the thermal Hall conductivity were measured by using the steady state method at low temperatures down to 0.3 K and in magnetic fields up to 14 T.[32-34] Heat current was generated by chip resistance and the thermal gradients were measured by using three resistor thermometers. The direction of heat current was along the (111) plane while the magnetic fields were applied along the [111] axis. The specific heat was measured by the relaxation method in the temperature range from 2 to 300 K using a commercial physical property measurement system (PPMS, Quantum Design).

## 3. Results and discussions

Figure 1 shows the zero-field specific heat of $Pr_2Zr_2O_7$ single crystal. The temperature dependence of the specific heat $C_p(T)$ does not exhibit sharp features that would result from long-range order transition. Besides, there is a large broad peak at low temperatures, indicating significant magnetic contribution from short-range correlation or spin fluctuations. In addition to the magnetic specific heat, $C_m$, the total specific heat, $C_p$, in the insulating $Pr_2Zr_2O_7$ should include: the lattice specific heat, $C_{ph}$, the crystal electric field (CEF) contribution, $C_{CEF}$, and the nuclear contribution, $C_N$. The nuclear Schottky specific heat from $Pr^{3+}$ nuclear moment is negligibly small in the present temperature range. Apparently, the phonon specific heat is dominant in the high-temperature regime from several tens of kelvins to 300 K. However, it was found that the standard Debye formula cannot fit the high-temperature data. One known reason for the deviations of high-temperature specific heat from the Debye model is the contribution of optical phonons, which can be described by the Einstein model.[34-38] The phonon spectra of $Pr_2Zr_2O_7$ should consist of 3 acoustic branches and 30 optical branches. We found that the high-temperature data can be fitted by the formula

$$C_{ph}(T) = 3N_D R \left(\frac{T}{\theta_D}\right)^3 \int_0^{\theta_D/T} \frac{x^4 e^x}{(e^x-1)^2} dx + 3N_{E1} R(\theta_{E1}/T)^2 \frac{\exp(\theta_{E1}/T)}{[\exp(\theta_{E1}/T)-1]^2} + 3N_{E2} R(\theta_{E2}/T)^2 \frac{\exp(\theta_{E2}/T)}{[\exp(\theta_{E2}/T)-1]^2} + 3N_{E3} R(\theta_{E3}/T)^2 \frac{\exp(\theta_{E3}/T)}{[\exp(\theta_{E3}/T)-1]^2}$$

where $x = \hbar\omega/k_B T$, $\omega$ is the phonon frequency, $k_B$ is the Boltzmann constant, and $R$ is the universal gas constant. Here, the first term is the contribution of three acoustic phonon branches using the Debye model ($N_D = 3$), while the other terms are the contributions from the optical branches using the Einstein model ($N_{E1} = 3$, $N_{E2} = 10$ and $N_{E3} = 17$). The other parameters are the Debye temperature, $\Theta_D = 405$ K, and three Einstein temperatures, $\Theta_{E1} = 95$ K, $\Theta_{E2} = 253$ K and $\Theta_{E3} = 575$ K. The obtained Debye temperature is comparable to that obtained by Kimura et al..[6] The fitting results are taken as the lattice specific heat of $Pr_2Zr_2O_7$.

$C_{CEF}$ was calculated by using a standard two-level Schottky formula

$$C_{\text{sch}}(T) = \frac{N\varepsilon_1^2}{k_B T^2} \frac{g_0}{g_1} \frac{\exp\left(\frac{\varepsilon_1}{k_B T}\right)}{[1 + \frac{g_0}{g_1}\exp\left(\frac{\varepsilon_1}{k_B T}\right)]^2}$$

where $N$ is the number of magnetic atoms per mole of molecule, $\varepsilon_1$ represents the energy gap between the ground state and first excited level, $g_0$ and $g_1$ represen the corresponding degeneracy of the ground state and the first excited state, respectively. The CEF energy levels of $Pr^{3+}$ have been obtained from inelastic neutron scattering.[6] Then, we can get the low-temperature magnetic specific heat by subtracting the phononic term and the CEF term from the raw data, as shown in Fig. 1(b). The broad peak of magnetic specific heat around 2 K should be related to the short-range magnetic correlation or spin fluctuations, which is a common behavior in the rare-earth pyrochlore materials. In particular, the broad peak at 2 K can be associated with the proliferation of thermally activated magnetic monopoles,[6] while the position of the peak is similar to that for $Dy_2Ti_2O_7$.[4] Another possibility is that the crystal defects produce a random transverse field acting on the non-Kramers $Pr^{3+}$ ions, which causes the CEF doublet state to split and leads to an additional CEF contribution at low tempeatures.[7,8]

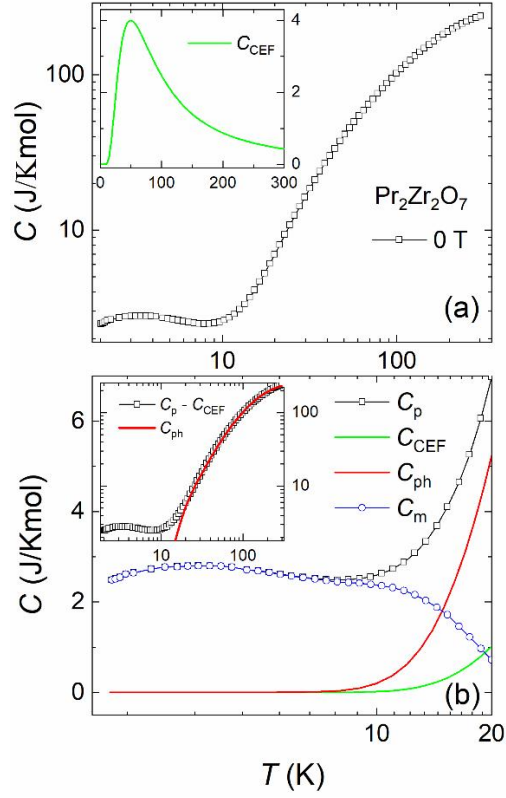

**Figure 1.** (a) Temperature dependence of the specific heat for Pr$_2$Zr$_2$O$_7$ single crystal in zero field. Inset: Temperature dependence of the $C_{CEF}$. (b) Magnetic specific heat $C_m$ after subtracting lattice and CEF contributions. Inset: Temperature dependence of the remaining specific heat after subtracting the CEF contribution at zero field. The red line is the fitting of phonon specific heat as described in the main text.

Figure 2 shows the temperature dependence of thermal conductivity $\kappa_{xx}$ in 0 and 14 T fields for Pr$_2$Zr$_2$O$_7$ single crystal. In zero field, the $\kappa_{xx}$ displays rather small values at low temperatures. We can calculate the mean free path of phonons by assuming phonon is the only heat carrier at low temperatures. The phononic thermal conductivity can be expressed by the kinetic formula $\kappa_{xx}^{ph} = \frac{1}{3}C_{ph}v_{ph}\ell_{ph}$,[39] where $C_{ph} = \beta T^3$ is phonon specific heat at low temperatures, $v_{ph}$ is the averaged velocity and $\ell_{ph}$ is the mean free path of phonons. Here, $\beta = 3.54$ J/m$^3$K$^4$ is obtained from the above fitting of the phononic specific heat, using the relation $\beta = \frac{12\pi^4}{5}\frac{RS}{\theta_D^3}$, where $R$ is the universal gas constant, $S$ is the number of atoms in the molecular formula, and $\Theta_D$ is the Debye temperature. From the Debye temperature $\Theta_D = 405$ K, we find $v_{ph} = 3255$ m/s, using the relation $v_{ph} = \frac{k_B T}{\hbar}T_D(6\pi^2 N/V)^{-1/3}$, where $N$ is the number of atoms per unit cell and $V$ is the unit cell volume. Then the mean free path of phonons can be obtained. In the boundary scattering limit, $\ell_{ph}$ is given by the effective width $W = 2\sqrt{wt/\pi}$, where $w$ and $t$ are width and thickness of the crystal. The inset of Fig. 2 shows the calculated $\ell/W$ at low temperatures for the 0 T and 14 T data. It is found that in zero field the $\ell_{ph}$ is very small; even at the lowest temperature of 0.3 K, $\ell$ is 2 orders of magnitude smaller than the $W$. In addition, the $\kappa_{xx}$ shows a $T^{1.24}$ dependence at very low temperatures, which is much weaker than the standard $T^3$ behavior of phonon thermal conductivity at the boundary scattering limit.[39] All these indicates that there is still some kind of microscopic scattering on phonons and the boundary scattering limit is not achieved at temperatures as low as 0.3 K.

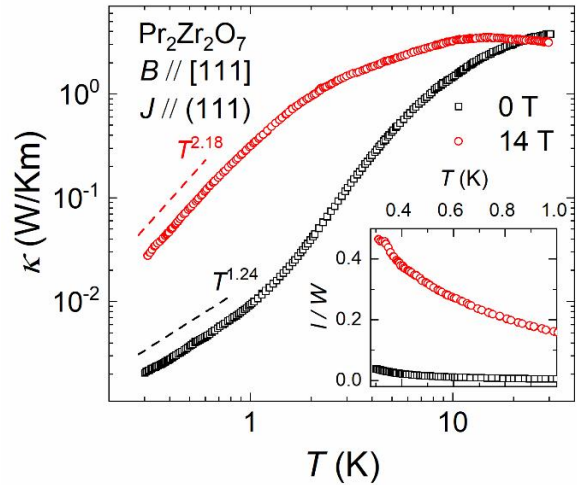

**Figure 2.** Temperature dependence of the thermal conductivity along the [111] axis of $Pr_2Zr_2O_7$ single crystal in 0 and 14 T magnetic fields. The dashed lines show the $T^{1.24}$ dependence and the $T^{2.18}$ dependence at very low temperatures. The inset shows the temperature dependence of the ratios of the phonon mean free path $\ell$ to the averaged sample width $W$. $\ell$ is calculated assuming that the $\kappa_{xx}$ is purely phononic.

The above calculation is based on the assumption of a purely phononic heat transport. If there were a magnetic term of heat transport, the phonon mean free path would be even smaller. Thus, it can be concluded that there are magnetic excitations at low temperatures, which can be spinons or magnetic monopoles, that scatter phonons rather strongly.[40] It should be pointed out that the magnetic-monopole excitations were believed to be negligible at very low temperatures because the energy barrier for flipping a spin or creating a pair of magnetic monopoles is about 1.6 K.[6] Unlike the classical case, the number of monopoles in the quantum spin ice do not disappear suddenly but decays exponentially with decreasing. Therefore, it is likely that the strong scattering at low temperatures is mainly caused by spinons. The $\kappa$ has a great enhancement at 14 T and the mean free path of phonons is also strongly increased, but it is still smaller than W at 0.3 K. In addition, the $\kappa$ shows a $T^{2.18}$ dependence at very low temperatures, which is, however, still much weaker than the $T^3$ behavior at the phonon boundary scattering limit. These indicate that 14 T field is not strong enough to completely suppress the magnetic scattering of phonons.

Figure 3 shows the field dependence of $\kappa_{xx}(B)$ for $B$ ∥ [111]. Below 1.4 K, there are three characteristic regimes: low-field regime where $\kappa_{xx}(B)$ increases slowly with increasing $B$, intermediate-field regime where $\kappa_{xx}(B)$ increases rapidly start from 5 T, and high-field regime where $\kappa_{xx}(B)$ exhibits a saturation behavior. Above 1.4 K, there are two characteristic regimes: low-field regime where $\kappa_{xx}(B)$ decreases with increasing B, and high-field regime where $\kappa_{xx}(B)$ increases rapidly which does not exhibit a saturation behavior. These data are rather comparable to those reported by Tokiwa et al..[41] In $Pr_2Zr_2O_7$, heat can be transferred by phonons and magnetic excitations, that is, $\kappa_{xx} = \kappa_{xx}^{ph} + \kappa_{xx}^{m}$. The previous magnetization data indicated that the spins are aligned rather quickly at low fields and the polarization is not achieved at ~ 10 T.[42-47] From above temperature dependence of $\kappa_{xx}$, it is known that in zero field there are significant magnetic scattering effects on phonons. Therefore, the quick increase of $\kappa_{xx}(B)$ at fields above 5 T and the tend to saturate in the high-field regime are likely due to the suppression of magnetic excitations and their scattering of phonons, associated with the spin polarization.

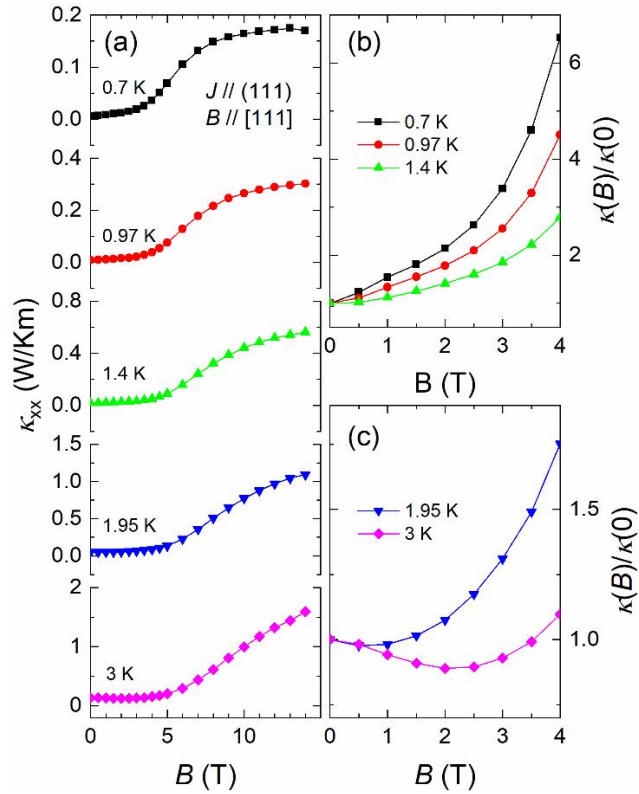

**Figure 3.** (a) Magnetic-field dependencies of the thermal conductivity of $Pr_2Zr_2O_7$ crystals at low temperatures. (b,c) Zoom-in of the low-field data at different temperatures.

In low-field regime, the $\kappa_{xx}(B)$ behaviors are not the same at different temperatures. An important information for the elementary excitations is provided by $\kappa_{xx}(B)$ in the low-field regime, where $\kappa_{xx}(B)$ increases slowly with increasing $B$ below 1.4 K (Fig. 3(b)). There are two competing impacts on thermal conductivity induced by the magnetic field, that is, the magnetic excitations can both carrying heat and scattering phonons. The zero-field thermal conductivity shows that at temperatures as low as 0.3 K, the phonon scattering is still very strong. The specific heat results show that the contribution of the CEF is negligibly small at low temperatures and is irrelevant to the phonon scattering. Therefore, the phonon scattering at low temperatures comes from low-energy magnetic excitations. As mentioned earlier, the magnetic monopole is suppressed by the energy gap at very low temperatures, and the magnetic excitations are spinons here. The present results are consistent with that reported by Tokiwa et al., which revealed a gapped behavior of the thermal conductivity at low-temperature regime (0.7–1.4 K).[41]

Above 1.4 K, $\kappa_{xx}(B)$ decreases with increasing $B$ in low fields. This low-field behavior of $\kappa_{xx}(B)$ is likely due to the field-suppression of some magnetic thermal transport because it cannot be explained by spin-phonon scattering effect, which always increases $\kappa_{xx}(B)$ with increasing $B$. And it also cannot be explained by the resonance scattering, which is due to a resonance between the Zeeman gap $g\mu_B H$ and the phonon, where $g$ is the $g$ factor and $\mu_B$ is the Bohr magneton. Because the suppression peak will occur when the Zeeman energy is equal to the peak of the Debye distribution function ($\sim 4k_B T$).[15] And our results are shown that the dips appear in different fields (Figure S6). Although we only have two pieces of data, we believe that it is not the result of phonon resonance scattering. So the magnetic monopoles are most likely to be excited in this case.[6,8,48] The decrease of $\kappa_{xx}(B)$ with $B$ implies that the number of magnetic monopoles is reduced with increasing $B$. This initial reduction is expected in the dispersionless classical magnetic monopoles with gap $2J_{zz}$ (1.6 K).[6] However, the number of magnetic monopoles will decay exponentially with decreasing temperature below 1.6 K. With the increasing temperature, the position of dip moves to higher field, which is a typical behavior of magnetic monopoles acting as heat carriers.[49]

Figure 4(a) shows the thermal Hall conductivity ($\kappa_{xy}$) as a function of the magnetic field for $Pr_2Zr_2O_7$ single crystal. Although we swept field up to 14 T, only the low-field data are of high quality. At high fields the transverse temperature gradient is too small to be properly measured, and it is not clear whether there is a thermal Hall effect in high fields or not (See Figure S7-S11 in the Supporting Information). At low temperatures, the $\kappa_{xy}$ displays a maximum at ~ 3 T. Figure 4(b) shows the Hall angle calculated by using the relation $\tan\theta_B = \kappa_{xy}/\kappa_{xx}$, which can represent the intrinsic thermal Hall response. And the thermal Hall response is divided into two special temperature regions for discussion. Above 1.4 K, the $\kappa_{xx}$ decreases with $B$ in low fields, which is similar to the $\kappa_{xx}(B)$ and indicates the presence of magnetic monopoles. At the same time, the maximum of the thermal Hall angle also moves towards high field as temperature increases. For example, the $\kappa_{xx}$ reaches a minimum near 1 T at 1.95 K and the thermal Hall angle reaches a maximum near 1.5 T. So, it is likely that the thermal Hall response above 1.4 K is caused by magnetic monopoles. Below 1.4 K, the $\kappa_{xx}$ shows a slow increase with $B$ at low fields, indicating that the magnetic field suppresses the spin fluctuations. At low temperatures, the spinons can both scatter phonons and carry heat. It is possible that the thermal Hall response below 1.4 K and at low fields is caused by spinons.

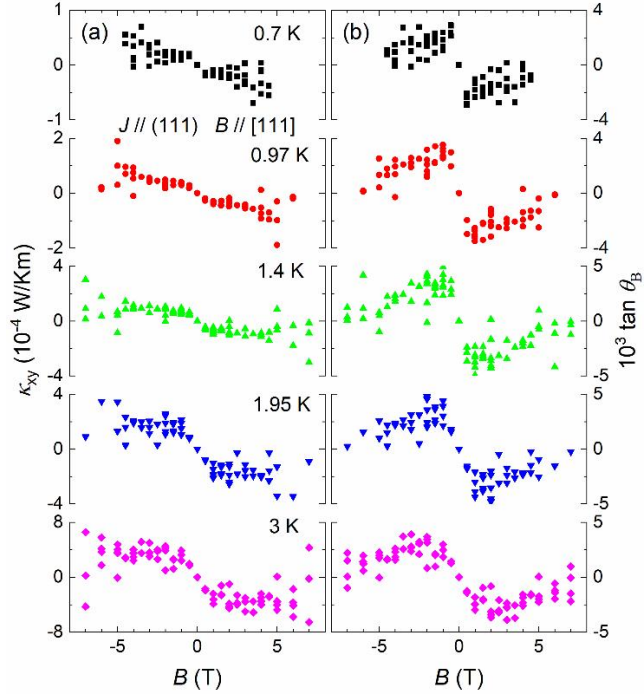

**Figure 4.** (a) Magnetic field dependencies of the thermal Hall effect of $Pr_2Zr_2O_7$ single crystal for $B \parallel [111]$. (b) Magnetic field dependencies of the Hall angle $\tan\theta_B$.

Can the $\kappa_{xy}$ of $Pr_2Zr_2O_7$ be caused by phonons? Firstly, the behavior thermal Hall angle provides an important information. If it were the contribution of phonons, the maximum of $\tan\theta_B$ would not move with changing temperature. Secondly, in the phonon scenario the skew scattering from local moments in the disordered state can yield a weak Hall signal at low $B$. As $B$ increases, alignment of the moments should lead to an increase of $\tan\theta_B$ until spin saturated.[50] Instead, the opposite is observed in the present work: $\tan\theta_B$ rapidly falls to nearly zero with increasing $B$. More importantly, with the rapid increase of $\kappa_{xx}(B)$ above 5 T the $\kappa_{xy}$ seems to vanish. In summary, the thermal Hall response is caused by magnetic excitations, including magnetic monopoles as well as spinons.

## 4. Conclusions

In conclusion, we have grown single crystals of $Pr_2Zr_2O_7$ and studied its specific heat, thermal conductivity, and thermal Hall conductivity at low temperatures. The main experimental results include: (i) The specific heat shows a broad peak at 2 K, which is likely caused by the short-range spin correlation or the inhomogeneous distribution of the ground-state doublet. (ii) The thermal conductivity shows an initial reduction with $B$ above 1.4 K, which can be attributed to emergent magnetic monopoles, and a slow increase below 1.4 K, which can be attributed to the phonon scattering caused by spinons. (iii) The thermal Hall conductivity shows a maximum at low field, which is caused by magnetic excitations including magnetic monopoles as well as spinons. It is a rare example of a thermal Hall response caused by spinons and magnetic monopoles.

## ASSOCIATED CONTENT

**Supporting Information**

The supporting information is available free of charge.

Additional material characterization data, sample measurement configuration (PDF).

## AUTHOR INFORMATION

**Corresponding Author**

*E-mail: wjchu@mail.ustc.edu.cn

## AUTHOR DECLARATIONS

**Conflict of Interest**

The authors have no conflicts to disclose.

**Author Contributions**

Wenjun Chu: Investigation; Experiment; Writing – original draft. Xuefeng Sun: Writing – review.

DATA AVAILABILITY

The data that support the findings of this study are available within the article and its supplementary material.

ACKNOWLEDGMENT

This work was supported by the National Natural Science Foundation of China (Grant Nos. U1832209 and 11874336).